# *cryptoRAN*: A review on cryptojacking and ransomware attacks w.r.t. banking industry - threats, challenges, & problems

Naresh Kshetri
CyROC, School of Business & Technology
Emporia State University
Emporia, Kansas, USA
nkshetri@emporia.edu

Mir Mehedi Rahman
CyROC, School of Business & Technology
Emporia State University
Emporia, Kansas, USA
mrahman2@g.emporia.edu

Sayed Abu Sayeed
CyROC, School of Business & Technology
Emporia State University
Emporia, Kansas, USA
ssayeed@g.emporia.edu

Irin Sultana
CyROC, School of Business & Technology
Emporia State University
Emporia, Kansas, USA
isultana@emporia.edu

*Abstract*— In the banking industry, ransomware is a well-known threat, but since the beginning of 2022, cryptojacking, an emerging threat is posing a considerable challenge to the banking industry. Ransomware has variants, and the attackers keep changing the nature of these variants. This review paper studies the complex background of these two threats and scrutinizes the actual challenges, and problems that the banking industry and financial institutions face. These threats, though distinct in nature, share commonalities, such as financial motivations and sophisticated techniques. We focus on examining the newly emerged variants of ransomware while we provide a comprehensive idea of cryptojacking and its nature. This paper involves a detailed breakdown of the specific threats posed by cryptojacking and ransomware. It explores the techniques cybercriminals use, the variabilities they look for, and the potential consequences for financial institutions and their customers. This paper also finds out how cybercriminals change their techniques following the security upgrades, and why financial firms including banks need to be proactive about cyber threats. Additionally, this paper reviews the background study of some existing papers, finds the research gaps that need to be addressed, and provides suggestions including a conclusion and future scope on those disputes. Lastly, we introduce a Digital Forensics and Incident Response (DFIR) approach for up-to-date cyber threat hunting processes for minimizing both cryptojacking and ransomware attacks in the banking industry.

## I. INTRODUCTION

In the past few years, cyber threats to the banking and financial businesses have grown a lot. As a result of how often they happen, cryptojacking and ransomware attacks have become major security worries. The goal of these attacks is to take advantage of weak spots in both traditional and digital banking systems. The amount of unlawful cryptocurrency mining has increased after 2019 [1]. This makes financial institutions' security, trustworthiness, and reputation very difficult. Random incidents of ransomware and illegal crypto mining are becoming more common, which has a big effect on the provided industry.

Up to the end of 2022, the most popular areas for cryptojacking were the government, healthcare, and education. However, there has been a "dramatic reshuffling" in these sectors [2]. The study claims that "cryptojacking targeting the financial industry skyrocketed to 269%" [2]. The second-highest number of attacks is on the retail sector; the number against the finance industry is currently five times higher [2].

Cryptojacking refers to the unauthorized utilization of computational resources and energy for the purpose of engaging in cryptocurrency mining, hence presenting a potential risk to the efficient operation of computer systems [3]. On the other hand, ransomware is a kind of harmful software that encrypts data so that users can't access it. In order to get the data back, victims are forced to pay an amount. The malevolent conduct results in the loss of information and disruptions to normal functioning [4].

Section II provides a comprehensive exposition of the background study conducted on current papers and incidents pertaining to threats in the financial and banking industry. Section III includes an analysis of the anticipated challenges, issues that banking sectors are likely to encounter in the anticipated future. Section IV examines the present circumstances, assessing its level of preparedness to confront difficulties and measures. Section V will examine the utilization of digital forensics and incident response (DFIR) methodologies in the context of cryptojacking and ransomware occurrences. Section VI will consist of a comprehensive conclusion and an exploration of future prospects with findings obtained and provide strategies for the future.

## II. BACKGROUND STUDY

This analysis will investigate the methodologies utilized in these attacks, the possible consequences that may arise, and the specific weaknesses that are targeted.

In [1], Varlioglu et al., (2022), As a first step, they read a lot of academic papers and business reports about the new "Fileless Cryptojacking" threat. Next, they presented a new threat-hunting-focused DFIR method with clear steps. According to Varlioglu et al. cryptojacking is a type of attack in which attackers utilize a victim's computer to mine cryptocurrency without their permission. Fileless malware, on the other hand, is not like traditional malware because it doesn't use executable files or scripts to spread. Fileless malware can use a variety of techniques to infect systems, such as exploiting vulnerabilities in legitimate software or injecting malicious code into memory. The combination of cryptojacking with fileless malware is very challenging to detect and delete.

In [5], Aldauiji et al., (2022), the authors gave a comprehensive analysis of Cyber Threat hunting tactics employed for the purpose of identifying ransomware attacks as well as providing an explanation of the Cyber Threat Intelligence (CTI) technique. People get ransomware when it encrypts their files and asks for money to get the decryption key. These cyberattacks create significant financial losses and disruption to operations. Cybercriminals are using more advanced methods to make ransomware. They target more groups, such as businesses, governments, and people. Cyber threat hunting (CTH) is an effective way to find malware and other types of cyber threats by constantly looking for strange activity on a system or network. CTH helps to find both known and new threats. Modern CTH systems use internal data sources and reactive methods to find strange behavior. So, ransomware attacks that are brand new or very well hidden are very difficult to detect by CTH.

In [6], Hasan and Al-Ramadan, (2022), were attempting to reveal how well-prepared Iraqi private banks were for cyberattacks. The study found that customers are well aware of cyberattacks. A new study from the Iraqi Banking Association says that even during strong attacks, Iraqi private banks keep a certain amount of security. Some banks are still hesitant to offer online banking, though, because they are worried about the safety of public internet access. But it depends on much money the bank has and how well it knows how to keep customer information and internal system assets safe from hackers.

In [7], Balbaa et al., (2022), they looked at how the war between Russia and Ukraine affected economies around the world. According to the article, the invasion creates disruptions in the international supply chain, causing energy prices to soar, and pushing up inflation. Since Russia and Ukraine are major exporters of oil, gas, and wheat, the war has also affected the energy and food sectors. The conflict shows the interdependence of the global economy among countries. The economic impacts of the conflict reach even those countries that are not directly involved in the conflict. Additionally, the conflict highlights the importance of energy security and supply chain diversification. Political leaders should use negotiations to resolve conflicts between Ukraine and Russia.

In [8], Zimba, (2022), it presented This article used Bayesian Attack-Network modeling to model crimeware-driven financial cyberattacks using GameOver Zeus malware. For conditional probabilities, the model uses Common Vulnerabilities and Exposures (CVEs). It shows how weaknesses can be used improperly or easily gotten around. Although the model shows that the GameOver Zeus botnet is complicated, it is hard to stop these kinds of attacks because infected hosts can live at any level. The authors recommend that law enforcement agencies collaborate to target all layers of the botnet, including the C&C, Proxy, and P2P layers. In order to protect themselves from cyberattacks using this attack model, they also advise businesses to focus on protecting nodes with the maximum vertex degree and clustering coefficient.

### III. CRYPTOJACKING AND RANSOMWARE ATTACKS

The impact of cryptojacking on individuals may not manifest immediately; nonetheless, it does result in hardware malfunctions and grants unauthorized access to cybercriminals to victims' computer systems [9]. Moreover, in the event that a corporation experiences several instances of computer infection, it may incur significant financial losses due to increased electricity consumption and the necessity to replace strained processing equipment [9].

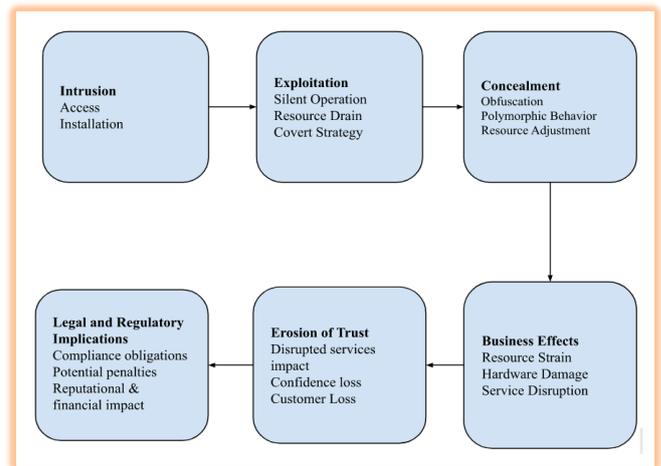

Figure 1: Cryptojacking lifecycle in banking and financial institutions

For several years, over 80% of businesses have faced increased cyberattacks, creating challenges in cybersecurity planning [10]. Ransomware attacks, targeting both individuals and corporations online, compel computer owners to pay due to their sophistication and widespread infection, resulting in substantial global losses annually [10]. Reported in over 150 countries, these attacks see an annual rise in average ransom payouts, varying from $1,000 to thousands, with demands exceeding $1 million commonly impacting larger enterprises [10].

Cryptojacking quietly operates in the background, evolving from browser-based attacks to focus on critical servers through file-based intrusions, presenting itself in three ways: exclusively in RAM using malicious scripts, within websites, and in operating systems and ROM [1]. Memory-centric crypto-hijacking operates stealthily within the system, employing tools like PowerShell or Windows Management Instrumentation (WMI) alongside robust registry-based techniques to avoid detection, making it more perilous than typical browser or host-based cryptojacking [1]. In 2018, a security flaw in the Browsealoud plugin developed by Texthelp allowed cryptocurrency to be mined covertly across a wide variety of websites by taking advantage of visitors' CPUs [11]. Without the users' knowledge, this JavaScript-based assault drained their CPU resources. Unknown actors exploited the plugin's code to inject Coinhive's controversial JavaScript-based Monero miner, leading to the illegal mining of cryptocurrency on the compromised sites [12]. Utilizing stealthy malware known as host-based cryptojacking, hackers obtain access to the resources of the victim host and transform it into a zombie machine for their own malicious purposes [12]. Host-based malware necessitates installation on the host system, in contrast to in-browser cryptojacking malware that gains access to use a web app to target the sufferer's computer power. Consequently, these threats are commonly introduced to the target system via malware in drive-by download schemes, vulnerabilities, integration into third-party applications, and social engineering methods [12].

Ransomware attacks are getting smarter and using more than one method. Key-chain and file-locker malware are the two main types. Crypto-ransomware is when an attacker locks up a victim's important files by encrypting them with strong algorithms like RSA or AES and demanding a payment [13]. Locker ransomware, on the other hand, locks the victim's machine and asks for money to unlock it, but it doesn't encrypt the data [13]. Cyber defense protects against evolving ransomware by using strong security measures. It blocks different ransomware types, like crypto-ransomware and locker ransomware, using real-time checks and stopping powerful encryption methods [14]. By focusing on prevention and quick responses, cyber defense keeps systems safe from ransomware attacks, securing sensitive data and ensuring resilience against these threats [14].

Websites using a plugin faced a security breach, leading to secret cryptocurrency mining on visitors' computers [15]. This exposed the plugin's vulnerability, affecting many global sites, including government and educational ones. It shows how one compromised plugin can impact many websites [15]. After a short mining period, quick actions were taken to stop the plugin and involve authorities with a proposed solution. This event stresses the need for better security when dealing with outside plugin content [15]. According to the Federal Reserve Board's research, ransomware poses a serious risk to the US economy, especially with regard to financial institutions. Substantial threats arise from sophisticated DDoS attacks and ransomware-as-a-service (RaaS), which cause more disruptions [16]. Financial services saw a nearly twofold increase in ransomware occurrences between 2021 and 2023, with 81% of organizations experiencing data encryption [16]. Furthermore, 254 million records were exposed by 566 breaches in 2022, highlighting the growing threat posed by data breaches that mostly affect institutions in the United States, Argentina, Brazil, and China [16]. These occurrences highlight how important it is to have strong cybersecurity safeguards in place, especially considering that hacks in the finance sector cost $5.9 million [16].

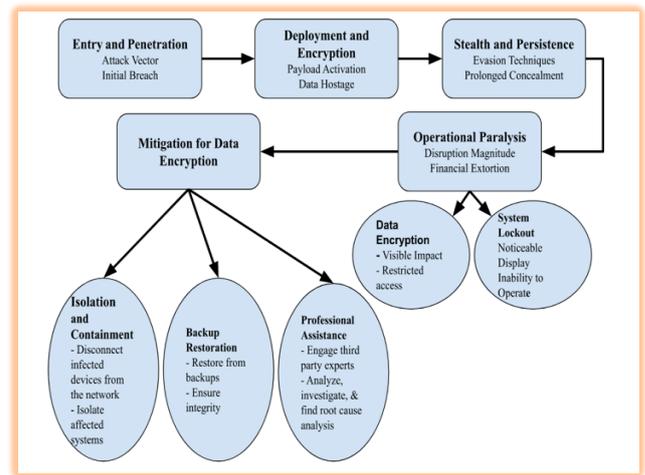

Figure 2: Ransomware lifecycle in banking and financial institution

## IV. CHALLENGES AND PROBLEMS AHEAD

Cryptojacking and ransomware create a wide range of difficulties and have far-reaching consequences in the banking sector. The banking industry worldwide is primarily influenced by two key factors. The first factor is the use of diverse protocols with standardization for communication among customized devices, resulting in heterogeneous data. The second factor is the influx of a large volume of events into data and communication networks, encompassing a wide array of information and occurrences [17]. Information and Communication Technology (ICT) is changing quickly in the cyber age. This has caused many privacy and security problems, which is why cybersecurity has become such an important area [17].

The goal of the increased cooperation between fintech companies and traditional financial institutions is to improve digital infrastructure, agility, innovation, and service quality [18]. This partnership, however, also prompts worries about heightened vulnerability to cybersecurity risks, including dangers like malware assaults, data breaches, integrity problems from fintech companies, and an increase in financial institution cyberattacks [18]. Problems such as unencrypted data, altered data, untrustworthy third-party services, and spoofing present significant hazards that can lead to monetary losses and damage to one's reputation [19]. To protect people's financial security, it is imperative that these issues be addressed jointly, which calls for a concerted effort to strengthen cybersecurity safeguards. Solutions including device identification, secure browser protection, one-time password (OTP) tokens, and transactional monitoring can be implemented to mitigate potential vulnerabilities detailed in the study [19] and improve the security of digital banking platforms.

Bank security risks have increased as a result of the transition to remote labor, particularly during the COVID-19 epidemic, as a result of shortened workweeks and a greater reliance on remote access, which exposes vulnerabilities [20]. Despite precautions, identifying insider threats in remote employment is still difficult [20]. Proactively addressing developing vulnerabilities requires real-time monitoring, updated controls, alarm systems, and anticipating of dangers connected to cryptography [20]. Technological developments necessitate effective defenses against hacker attacks, particularly with regard to cryptocurrencies and blockchain, which highlights the importance of close observation within the digital banking ecosystem [20]. Banks that use artificial intelligence (AI) and technological solutions for digitalization must always use the most recent security measures [20]. The enduring nature of cyber threats highlights the pressing requirement for strong security protocols that address ransomware, phishing, insider threats, hacking, and data theft. Implementing strong security protocols, including endpoint protection, VPN solutions, and network security measures, is crucial for safeguarding business infrastructure in remote work setups. Encouraging employees to seek IT support for technical concerns and ensuring secure communication tools are used are additional vital steps in maintaining a secure remote working environment [21].

Although cryptojacking may not seem as dangerous, it can have unexpectedly large and negative effects on banks [22]. Due to this, banks unwittingly pay for the high processing power required for mining cryptocurrencies, which raises electricity costs and deteriorates hardware components [22]. Since these attacks are often undetected due to their stealthy nature, it is challenging to determine their true costs. Furthermore, cryptojacking uses a lot of bandwidth, which hinders regular banking operations and impairs user performance [22]. This poses a severe risk to banks, raising the possibility of data breaches and perhaps slowing down operations [23]. The reputation and stability of a bank can be severely damaged by the ensuing financial losses and the costs of combating these attacks [24]. Cryptojacking, exemplified by Coin Hive's stealthy code diversion for cryptocurrency mining, not only disrupts computer processing but also poses severe risks to banks, potentially leading to data breaches, hindering operations, and causing financial losses, thereby damaging a bank's stability and reputation [25].

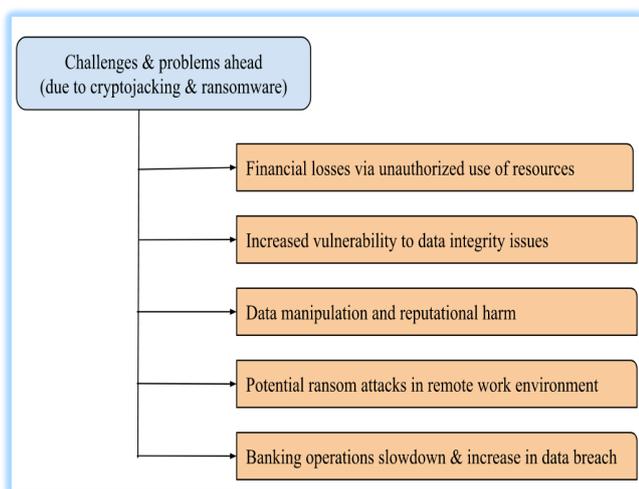

Figure 3: Challenges and problems ahead due to cryptojacking and ransomware attacks

## V. BANKING INDUSTRY AND SECURITY

A multifaceted approach is required for effective cybersecurity awareness in the banking sector, with a focus on top management commitment and comprehensive awareness programs across industries [26]. Sufficient funding allocation and managerial assistance are essential for the effective implementation of cybersecurity policies and the reduction of information security risks stemming from employee ignorance [26]. Developing an organizational culture that is robust to cybersecurity threats is essential for encouraging workers to follow security procedures, closing knowledge gaps, and creating a culture where security is the top priority [26]. Strong security measures are required for banks since they face a variety of cybersecurity risks, such as phishing, cyberterrorism, and identity theft [27]. In order to reduce these risks, it is critical to execute routine malware checks, encrypt data, ensure secure data transmission, employ biometrics and multi-factor authentication for account security, and increase user knowledge of phishing and fraudulent websites [27]. The 21st century's rapid technological advancements present both benefits and difficulties, underscoring the growing need to protect data from theft or misuse [28]. The need to adapt and develop a secure information environment for the future is highlighted by the emergence of new dangers like information wars and hacking as the Internet and related technologies grow [28]. To protect against increasing cyber threats, it is critical to work with cybersecurity professionals, invest in state-of-the-art solutions, and keep up with technical breakthroughs [27] [28].

In today's information-centric environment, establishing robust information protection systems is crucial given the escalating risks [29]. Achieving a balance between cost-effectiveness and security is essential to mitigate potential interruptions that could cause socio-economic instability in the realm of digital banking [29][30]. Various evaluation techniques, such as fuzzy logic, ranking systems, and statistical analysis, provide valuable insights into information security [29]. Open banking technologies like electronic money and online services are indispensable for updating services, meeting client demands, and ensuring enhanced data security for safe and accessible financial transactions [29]. Innovations in IT security and Big Data technologies play a pivotal role in reinforcing information security within the banking industry by safeguarding private information, addressing cyber threats, and managing substantial data volumes [29]. In the domain of digital banking, confidentiality, availability, and integrity act as crucial pillars, ensuring the accuracy, secrecy, and accessibility of critical information [30]. However, barriers to consumer confidence impede the adoption of state-of-the-art technology, impacting online transactions [30]. Trust is closely linked to the perception of security, influencing user risk perception regarding data availability, integrity, and confidentiality [30]. Proficiency in these elements is vital to enhancing trust and combating cyber threats in electronic banking. Notably, the CIA framework plays a pivotal role in fostering confidence and safeguarding confidential information, particularly in the changing world of financial technology [30].

It is absolutely necessary to protect financial transactions and data against threats such as cryptojacking and ransomware. Continuous education and culture with a strong emphasis on cybersecurity are two ways in which the industry can improve its level of security. The utilization of cutting-edge technology such as encryption and biometric verification in combination with interactive training for all staff levels helps to guarantee the security of customer accounts. Regular audits carried out by cybersecurity specialists are an important component in protecting against cyberattacks. Customers are more likely to place their faith in a company that places a premium on the privacy and dependability of their data. Methods that are flexible in terms of cybersecurity can keep up with the ever-evolving dangers, which will secure the future of digital banking as there is global rise in cybercrime, cyber defense, and countermeasures [31]. Protecting financial institutions against cryptojacking, ransomware, and other constantly morphing threats requires both preventative measures and a robust culture of cybersecurity.

## VI. DFIR APPROACH FOR CYBER THREATS

We can use digital forensics for investigating crimes, internal policy violations and it is the application of science to identify, collect, examine, & data analysis [32][33]. Like the same way in which computers are used for committing crimes, now computers are used by law enforcement to counter the crime. Considered both a science and an art, there is no deterministic procedure in digital forensics to answer directly. As long as it is associated with electronic crime (e-crime), digital evidence can be used to identify upcoming threats (including the banking industry) and relationships with other suspects. Despite several challenges in handling DFIR [34], it relies on self-defined best practices and experiences.

One popular cyber tool for data breaches, and fraud users when it comes to banking (e-banking) is phishing attacks. Most websites related to phishing are designed for Internet banking where attackers can acquire the financial information of the customers [35]. For such growing cybercrime, phishing attacks, and cyber fraud, we need some proper methodology to gather digital evidence that involves network traffic analysis for detecting intrusions [36]. Digital space (including recent technology like blockchain for financial institutions) has impacted information, trade, communication, and almost all industries, large/small businesses, supply chain management, government firms including online banking/e-banking [37]. Cyberspace is used for many illegal activities and any information can be changed by changing its statistical quantities. This information can then be used for more illegal activities [38].

When several other methods are inefficient in combating cyberattacks including cryptojacking and ransomware in the banking industry, digital forensics can be the key to locating and countermeasures attackers. Data in banks is a major target and it is one of the critical challenges to secure banking data [39]. Cybercrime includes dealings between people who are victims of crime, criminals who are cyber criminals, and criminals who are financial institutions. These deals are made through fake emails, social media, and other social engineering techniques [40]. Another approach of DFIR/digital forensics is that it can aid both mobile banking and branchless banking in such areas where banks (bank branches or auto-telling machines) cannot operate due to infrastructure problems [41].

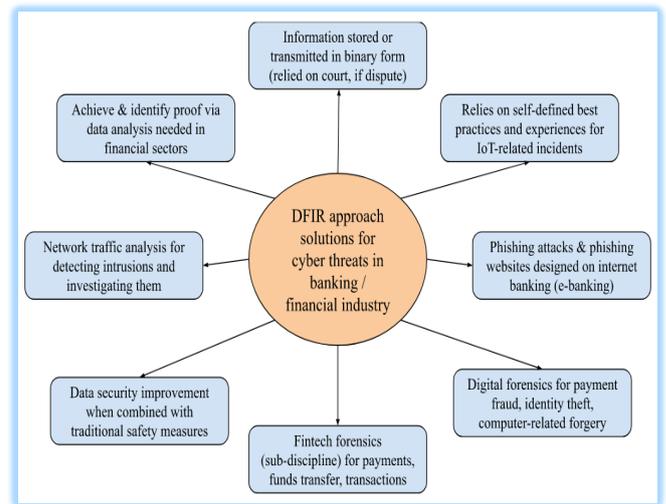

Figure 4: Digital forensics and Incident response (DFIR) approach solutions for ongoing (and future) cyber threats in the banking & financial industry

## VII. CONCLUSION & FUTURE SCOPE

Our work involved a detailed review of specific threats posed by cryptojacking and ransomware. We introduced the Digital Forensics and Incident Response (DFIR) approach for an up-to-date cyber threat hunting process for minimizing both cryptojacking and ransomware attacks. The challenges and problems ahead are discussed and reviewed in the scenario of the banking companies, financial institutions, and the security of banking customers. The challenges and problems pointed out can be used and helpful to minimize the cryptojacking and ransomware attacks and secure the systems, people, networks, and nation states. DFIR approach can act as a novel and effective countermeasure/counterintelligence in the prevention or slowdown of increasing cyber threats, cybercrimes and cyberattacks as well as recovering from data breach. Digital forensics can be sought as a solution for cyber threats, cyber-attacks in banking industry and financial industry for payments, funds transfer, and financial transactions etc. The future scope of the study can be extended to other types of cyber-attacks besides cryptojacking attacks and ransomware attacks. Security algorithms can be proposed with the help of DFIR approach (like cyber intrusion and evidence, breach response, post-mortem analysis) for identifying and investigating cyber incidents. We can also go beyond the banking industry and also include several other areas like the healthcare industry, automation, and higher education sectors to identify and mitigate cyber-attacks.